\documentclass[12pt,draft]{article}
\begin{document}
\title{Superfluid Phases of $^3$He in Aerogel.}
\author{I.A.Fomin\\
P. L. Kapitza Institute for Physical Problems, \\
ul. Kosygina 2, 119334 Moscow,Russia}
\date{ }

\maketitle
\begin{abstract}

Within the phenomenological approach a criterion for a choice of
the order parameter of the superfluid phases of $^3$He in aerogel in a
vicinity of the transition temperature is derived. The order parameter
of bulk B-phase of $^3$He meets this criterion and that of the bulk
A-phase doesn`t.  A class of order parameters of Equal Spin Pairing
(ESP) type meeting the derived criterion is discussed. Order parameters
belonging to this class are proposed as candidates for the observed
A-like superfluid phase of liquid $^3$He in aerogel. Effect of magnetic
field on the order parameters of this class is considered.

\end{abstract}


\section {Introduction}
Among the other physical objects with unconventional Cooper pairing
 the superfluid phases of $^3$He are best understood.
The pairing is referred as unconventional when simultaneously with
the gauge symmetry some other symmetries of a normal (nonsuperconducting or
nonsuperfluid) phase are broken. Liquid $^3$He, filling empty space in a high
porosity aerogel can be superfluid \cite{parpia}. The aerogel in that
case plays a role of impurities. This system provides a good possibility to
investigate effect of impurities on the superfluidity of $^3$He and more
generally on unconventional Cooper pairing. This knowledge can be applied to
metallic unconventional superconductors  UPt$_3$, UBe$_{13}$, Sr$_2$Ru$_2$O$_4$, UGe$_2$
etc., where different types of impurities are inevitably present. Aerogel is
different  from usually considered point-like impurities. It can be modelled
as a rigid frame consisting of randomly situated and randomly oriented strands
of $\approx 30 $\AA
thickness. For 98\% aerogel an estimated average distance
between the strands $\approx 200$\AA  is of the order of the superfluid
correlation length  $\xi_0$, which varies in $^3$He depending on a pressure
in the interval $160 - 500$\AA.  Estimated mean free path
$l\sim 1500 - 1800 $\AA is greater then $\xi_0$.  According to the theory
of superconducting alloys \cite{abri} at the unconventional Cooper pairing
even nonmagnetic impurities lower a temperature of transition in the superfluid
(or superconducting) state $T_c$ to the extent of a ratio
$\xi_0/l$ \cite{Larkin}.  Such suppression is observed in $^3$He.
Below $T_c$ in a system $^3$He+aerogel two superfluid phases are observed
\cite{osher}. By the analogy with the bulk $^3$He one of the observed phases
is referred as A-like phase and the other -- as B-like. Recent observation of
homogeneously precessing domain in the B-like phase \cite{dmit} is a very
strong evidence that the order parameter of the B-like phase is close to that
of BW-phase of the bulk $^3$He. Situation with the A-like phase is not so
clear. Identification of that phase rises important question, whether in a
presence of aerogel new superfluid phases of $^3$He can realize, with the
order parameter which is different from that of possible superfluid
phases of bulk $^3$He.  Answer to this question is a main objective of the
present paper. In what follows a procedure for determining the order
parameters of superfluid phases $^3$He in aerogel in a vicinity of $T_c$ is
formulated and applied to equal spin pairing (ESP) states
which are good candidates for description of the A-like phase.

\section {Interaction of aerogel with the order parameter}
At Cooper pairing with $l\ne 0$ except for the above mentioned general
suppression of $T_c$  effect of fluctuations in positions and orientations of
the strands of aerogel is of importance. In a vicinity of
$T_c$ this effect can be described phenomenologically by a random field, acting
on the order parameter. Corresponding term in the free energy can be constructed
using symmetry argument.
In $^3$He Copper pairs are formed in a state with the orbital moment $l=1$
and spin $s=1$. In that case the order parameter is a 3$\times$3
complex matrix $A_{\mu j}$, index
  $\mu$ refers to spin and  $j$ to orbital projection.
Interaction of  $^3$He with the strands of aerogel is due to the scattering of
quasiparticles on the strands. Quasiparticles change their momenta and interact
directly with the orbital part of the order parameter.
 Aerogel consists of magnetically inert SiO$_2$, but when immersed in the
liquid $^3$He it forms a layer of "solid" $^3$He
 on the strands. Quasiparticles of the liquid can exchange their spins
with that of atoms in the "solid" layer of $^3$He. This exchange provides
a mechanism of interaction of aerogel with the spin part of the order
parameter.  In experiments this mechanism can be switched off by addition of
$^4$He in a cell. Atoms of $^4$He substitute $^3$He in the "solid" layer and
make exchange impossible. Admixture of $^4$He  strongly influences magnetic
properties of superfluid $^3$He in aerogel. In what follows it is assumed that
the strands of aerogel are covered by a layer of $^4$He. In that case aerogel
interacts only with the orbital part of the order parameter $A_{\mu j}$ and in
a principal order on $A_{\mu j}$ a corresponding contribution to the free energy
has the form \cite{fom}:
$$
F_{\eta}=N(0)\int \eta_{jl}({\bf r})A_{\mu j}A_{\mu l}^*d^3r , \eqno(1)
$$
where $N(0)$ is the density of states at the Fermi level and
$\eta_{jl}({\bf r})$ -- a random static tensor field.
On the strength of  $t\to -t$ invariance tensor $\eta_{jl}({\bf r})$
is real and symmetrical, its isotropic part
$\frac{1}{3}\eta_{ll}({\bf r})\delta_{jl}$ describes  local variations of
$T_c=T_c({\bf r})$ due to fluctuations of the density of scatterers.
Anisotropic part
$\eta_{jl}({\bf r})-\frac{1}{3}\eta_{ll}({\bf r})\delta_{jl}
\equiv\eta^{(a)}_{jl}$ describes local splitting of  $T_c$ for different
projections of angular momenta because of the breaking of spherical symmetry by
the aerogel strands. Isotropic part of the random field will be included in
 $T_c=T_c({\bf r})$.
Absolute value of the random field can be estimated with the use of the results
of the paper \cite{Rainer}. By the order of magnitude it is
$|\eta_{jl}|\sim(x\xi_0/R)\sim(\xi_0/l)$, where $l$ -- is a mean free path,
$R$-- average radius of a strand $x$ -- fraction of a volume, occupied by
aerogel.  For a 98\% aerogel $\xi_0/l\sim 1/10$.
The field $\eta_{jl}({\bf r})$ varies on a scale of the average distance
between the strands $d\sim R/\sqrt{x}$. In a 98\% aerogel it is of the order of
 $\xi_0$. If the order parameter varies  on a scale of  $\sim d$ a
 loss of the gradient energy $\sim(\xi_0/d)^2$ exceeds a gain of energy in the
field  $\eta_{jl}({\bf r})$ to an extent of  $\frac{\xi_0}{R}\gg 1$.
For that reason the order parameter does not follow variations of the field
and does not form states localized on a scale  $\sim d$. It is possible
thou to form localized states with a scale $L\gg d$ even in a weak
 field $\eta_{jl}({\bf r})$ because of a degeneracy of the average order
 parameter $\bar A_{\mu j}$ over rotations in the orbital (momentum) space.
 According to Imry and Ma \cite{imry} a random field can destroy a long range
 order if the ordering is characterised by a continuously degenerate order
 parameter. They show in particular that for a vector order parameter
 ${\bf s}({\bf r})$ interacting with a random field ${\bf h}({\bf r})$ as:
$$
F_{IM}=-\int{\bf s}({\bf r})\cdot{\bf h}({\bf r})d^3r,          \eqno(2)
$$
long range order is destroyed by arbitrary small field.
The argument is following. The average value of the random
field ${\bf h}({\bf r})$ is
zero, i.e. $\frac{1}{L^3}\int{\bf h}({\bf r})d^3r\to 0$ at
$L\to\infty$, where $L$--is a linear scale of a region of integration.
The integral tends to zero as $\sim (d/L)^{3/2}$.
As the same power of $L$ tends to zero a gain of energy of the order
parameter due to its orientation in a field averaged over a region with a scale
$\sim L$. The loss of gradient energy decreases faster  $\sim (\xi_0/L)^2$ and
for large  $L$  formation of domains becomes advantageous. As a result the long
range order is destroyed.
General argument of Imry and Ma does not apply directly to the superfluid
$^3$He in aerogel.  The interaction $F_{\eta}$  differs from
$F_{IM}$ in a sense  that there exist finite $\bar A_{\mu j}$ for which
$F_{\eta}$ is zero for all possible $\eta^{(a)}_{jl}$. These $\bar A_{\mu j}$
 can be found from the equation
$$
\eta^{(a)}_{j l}\bar A_{\mu j}\bar A_{\mu l}^*=0,                   \eqno(3)
$$
or, equivalently from the equation which does not contain $\eta_{jl}$:
$$
\bar A_{\mu l}\bar A_{\mu j}^* +\bar A_{\mu j}\bar A_{\mu l}^*
 =\delta_{jl}\cdot const. \eqno(4)
$$
This equation determines the real part of a product
 $\bar A_{\mu j}\bar A_{\mu l}^*$,
its imaginary part is an arbitrary antisymmetric tensor.
When interaction with a random field turns to zero local reorientation of the
order parameter does not give a gain in the energy and  a long range order
is preserved. From that qualitative argument one concludes that
in a situation when average value of the order parameter gives a principal
 contribution to free energy eq. (3) gives a necessary condition of stability
 of a  phase corresponding to $\bar A_{\mu j}$  with respect to a random field
$\eta_{jl}({\bf r})$ \cite{fom_A}. In the next section
a procedure of finding of the order parameter $\bar A_{\mu j}$ in a presence
of a random field $\eta_{jl}({\bf r})$ is formulated. It will be shown that
the criterion (3) or (4) emerges as a natural requirement of consistency of the
procedure of minimization of the corresponding free energy.

\section {Selection of superfluid phases}
With the account of interaction eq.(1) Ginzburg-Landau functional takes
the following form:
$$
 F_{GL}=N(0)\int d^3r[\tau A_{\mu j}A_{\mu j}^*+
\eta_{jl}({\bf r})A_{\mu j}A_{\mu l}^*+\frac{1}{2}\sum_{s=1}^5 \beta_sI_s+
$$
$$
\frac{1}{2}\left(K_1\frac{\partial A_{\mu l}}{\partial x_j}
\frac{\partial A^*_{\mu l}}{\partial x_j}+
K_2\frac{\partial A_{\mu l}}{\partial x_j}\frac{\partial A^*_{\mu j}}{\partial
x_l}+K_3\frac{\partial A_{\mu j}}{\partial x_j}\frac{\partial A^*_{\mu
l}}{\partial x_l}\right)\Bigr] ,           \eqno(5)
$$
where $\tau=(T-T_c)/T_c$, $I_s$ -  4-th order invariants in the expansion of
the free energy over $A_{\mu j}$, we don`t need their explicit expressions (cf.
\cite{vollh}). Coefficients $\beta_1,...\beta_5,
K_1,K_2,K_3$ -- phenomenological constants.
In what follows we use for their evaluation the weak coupling values, in
particular we assume $K_1=K_2=K_3\equiv K$.  The gradient terms can acquire
random corrections as well e.g. of a form:  $u_j({\bf r})A_{\mu
l}\frac{\partial A^*_{\mu l}}{\partial x_j}$, where $u_j({\bf r})$ is a random
vector
\footnote{V.I.Marchenko pointed out to me a possibility of their
existence}.
Up to a prefactor $\hbar/m$ this is a local random velocity. The
terms of such form are obtained by extension of derivatives
$\frac{\partial}{\partial x_j}\to\frac{\partial}{\partial x_j}+u_j({\bf r})$ in
the expression for free energy (5). These terms do not influence the selection
of superfluid phases, that`s why they are not kept in what follows. Variation
 of the functional (5) over $A^*_{\mu j}$ renders an equation for the
 equilibrium order parameter:
 $$ \tau A_{\mu
 j}+\frac{1}{2}\sum_{s=1}^5 \beta_s\frac{\partial I_s}{\partial A^*_{\mu j}}-
\frac{1}{2}K\left(\frac{\partial^2 A_{\mu j}}{\partial x_l^2}+
2\frac{\partial^2 A_{\mu l}}{\partial x_l \partial x_j}\right)=
-A_{\mu l}\eta_{lj},                                 \eqno(6)
$$
 and variation over $A_{\mu j}$ -- the complex conjugated equation.

 Random field $\eta_{jl}({\bf r})$ according to the above estimation is small.
 Effect of a small random field on a conventional superconductor in a vicinity
 of $T_c$ was analysed by Larkin and Ovchinnikov \cite{LarkOv}. More
 complicated form of the order parameter in $^3$He and its degeneracy requires
 a nontrivial modification of that procedure.

Random field induces fluctuations of the order parameter $a_{\mu j}$
at its average value $\bar A_{\mu j}$,
i.e. $A_{\mu j}({\bf r})=\bar A_{\mu j}+a_{\mu j}({\bf r})$.
A long range order is established when
$\bar A_{\mu j}\ne 0$. This condition determines the transition temperature
$T_c$.  Not too close to $T_c$ fluctuation $a_{\mu j}$  is magnitude of the
first order on $\eta_{jl}$.  Let us expand eq.(6) at $A_{\mu j}=\bar A_{\mu j}$
keeping terms up to the second order on $a_{\mu j}$ and $\eta_{jl}$:
$$
\tau\bar A_{\mu j}+\tau
a_{\mu j}+\frac{1}{2}\sum_{s=1}^5 \beta_s\bigl[\frac{\partial I_s}{\partial
A^*_{\mu j}}+ \frac{\partial^2 I_s}{\partial A^*_{\mu j}\partial A_{\nu
n}}a_{\nu n}+ \frac{\partial^2 I_s}{\partial A^*_{\mu j}\partial A^*_{\nu
n}}a^*_{\nu n}+ $$ $$ \frac{1}{2}\left(\frac{\partial^3 I_s}{\partial A^*_{\mu
j}\partial A_{\nu n} \partial A_{\beta l}}a_{\nu n}a_{\beta l}+
2\frac{\partial^3 I_s}{\partial A^*_{\mu j}\partial A^*_{\nu n}
\partial A_{\beta l}}a^*_{\nu n}a_{\beta l}\right)-
$$
$$
\frac{1}{2}K\left(\frac{\partial^2 a_{\mu j}}{\partial x_l^2}+
2\frac{\partial^2 a_{\mu l}}{\partial x_l \partial x_j}\right)\Bigr]=
-\bar A_{\mu l}\eta_{lj}-a_{\mu l}\eta_{lj}.   \eqno(7)
$$
Now let us take average of the obtained equation over the length scales, which
are much greater then  the average distance between the strands of aerogel.
$$
\tau\bar A_{\mu
j}+\frac{1}{2}\sum_{s=1}^5 \beta_s\bigl[\frac{\partial I_s}{\partial A^*_{\mu
j}}+ \frac{1}{2}\bigl( \frac{\partial^3 I_s}{\partial A^*_{\mu j}\partial
A_{\nu n}\partial A_{\beta l}} <a_{\nu n}a_{\beta l}>+ $$ $$ 2\frac{\partial^3
I_s}{\partial A^*_{\mu j}\partial A^*_{\nu n} \partial A_{\beta l}}<a^*_{\nu
n}a_{\beta l}>\bigr)\bigr]= -<a_{\mu l}\eta_{lj}>.              \eqno(8)
$$
Except for $\bar A_{\mu j}$ the obtained equation  contains averages of
binary products of fluctuations $<a_{\nu n}a_{\beta l}>$ etc..
Equations for $a_{\mu j}$ are obtained by separation of fast varying terms
in eq.(7) and in its complex conjugated
$$
\tau a_{\mu
j}+\frac{1}{2}\sum_{s=1}^5\beta_s\bigl[ \frac{\partial^2 I_s}{\partial A^*_{\mu
j}\partial A_{\nu n}}a_{\nu n}+ \frac{\partial^2 I_s}{\partial A^*_{\mu
j}\partial A^*_{\nu n}}a^*_{\nu n}- $$ $$ \frac{1}{2}K\left(\frac{\partial^2
a_{\mu j}}{\partial x_l^2}+ 2\frac{\partial^2 a_{\mu l}}{\partial x_l \partial
x_j}\right)\bigr]= -\bar A_{\mu l}\eta_{lj},               \eqno(9)
$$

$$
\tau a^*_{\mu j}+\frac{1}{2}\sum_{s=1}^5\beta_s\bigl[
\frac{\partial^2 I_s}{\partial A_{\mu j}\partial A^*_{\nu n}}a^*_{\nu n}+
\frac{\partial^2 I_s}{\partial A_{\mu j}\partial A_{\nu n}}a_{\nu n}-
$$
$$
\frac{1}{2}K\left(\frac{\partial^2 a^*_{\mu j}}{\partial x_l^2}+
2\frac{\partial^2 a^*_{\mu l}}{\partial x_l \partial x_j}\right)\bigr]=
-\bar A^*_{\mu l}\eta_{lj},               \eqno(10)
$$

This is a linear inhomogeneous system of equations. As a consequence of
the discussed above degeneracy of the order parameter $\bar A_{\mu j}$
the corresponding homogeneous system has nontrivial solutions.
These are variations of $\bar A_{\mu j}$ and $\bar A^*_{\mu j}$ at an
infinitesimal rotation $\theta_n$:
$$
\omega_{\mu j}=\theta_ne^{jnr}\bar A_{\mu r},
\omega^*_{\mu j}=\theta_ne^{jnr}\bar A^*_{\mu r},         \eqno(11)
$$
where $e^{jnr}$ -- the absolutely anti-symmetric tensor.
It is convenient to use Fourier transformed equations (9),(10) for
$\eta_{jl}({\bf k})$ and $a_{\mu j}({\bf k})$.
For the further argument only a character of singularity of
$a_{\mu j}({\bf k})$ at $k\to 0$ is important. To find it we neglect
anisotropy of the gradient terms in eqns.(9),(10) and substitute
$\frac{1}{2}\bar K\left(\frac{\partial^2 a_{\mu j}}{\partial x_l^2}\right)$ and
$\frac{1}{2}\bar K\left(\frac{\partial^2 a^*_{\mu j}}{\partial x_l^2}\right)$
instead of
$\frac{1}{2}K\left(\frac{\partial^2 a_{\mu j}}{\partial x_l^2}+
2\frac{\partial^2 a_{\mu l}}{\partial x_l \partial x_j}\right)$ and
$\frac{1}{2}K\left(\frac{\partial^2 a^*_{\mu j}}{\partial x_l^2}+
2\frac{\partial^2 a^*_{\mu l}}{\partial x_l \partial x_j}\right)$.

Taking product of eq. (9) by  $\omega^*_{\mu j}$,  eq. (10) by
$\omega_{\mu j}$ and taking sum of both equations one arrives at the
following expression for the projection
$a_{\mu j}({\bf k})\omega^*_{\mu j}+a^*_{\mu j}({\bf k})\omega_{\mu j}
\equiv a^{\omega}({\bf k})$:
$$
a^{\omega}({\bf k})=-\frac{2}{k^2\bar K}
\left(\theta_n e^{jnr}Q_{rl}\eta^{(a)}_{lj}({\bf k})\right),     \eqno(12)
$$
where $Q_{rl}=\bar A_{\mu r}\bar A^*_{\mu l}+\bar A_{\mu l}\bar A^*_{\mu r}$ .
This projection gives a contribution to the averages
 $<a_{\nu n}a_{\beta l}>$  which is proportional to
$$
\int<a^{\omega}({\bf k})a^{\omega}(-{\bf k})>d^3k \sim
$$
$$
\theta_p e^{jpr}Q_{rl}\theta_q e^{mqs}Q_{sn}
\int<\eta^{(a)}_{lj}({\bf k})\eta^{(a)}_{mn}(-{\bf k})>\frac{d^3k}{k^4}.
$$
Form of the correlation function in the integrand after averaging over
direction of ${\bf k}$ is determined by the symmetry properties of
 $\eta^{(a)}_{lj}$:
$$
\int<\eta^{(a)}_{lj}({\bf k})\eta^{(a)}_{lj}(-{\bf k})>\frac{do}{4\pi}=
\Phi(k)[2\delta_{lj}\delta_{mn}-3(\delta_{lj}\delta_{mn}+
\delta_{lj}\delta_{mn})].                                         \eqno(13)
$$
Finally
$$
\int<a^{\omega}({\bf k})a^{\omega}(-{\bf k})>d^3k \sim
$$

$$
\theta_p\theta_q[e^{npr}e^{lqs}Q_{rl}Q_{sn}+(\delta_{pq}\delta_{rs}-
\delta_{ps}\delta_{qr})Q_{rn}Q_{sn}]\int\frac{\Phi(k)}{k^2}dk.   \eqno(14)
$$
The integral in the left-hand side diverges at the lower limit and
the averaged products $<a^{\omega}a^{\omega}>$ give diverging contributions
to eq.(5). To make the iteration scheme eqns.(5),(6) consistent
one has to require the coefficient in front of the diverging integral to be
zero.  Equating to zero the expression in square brackets in eq. (11) one
arrives at $Q_{rn}=q\delta_{rn}$, where $q$ is a real number. In terms of the
order parameter this condition reads as:
$$
\bar A_{\nu r}\bar A^*_{\nu
j}+\bar A_{\nu j}\bar A^*_{\nu r}= const\cdot\delta_{rj}.
$$
that coincides with the criterion (4).

The order parameters meeting criterion (4) can be refered as
"quasi isotropic" since the energy of their interaction with aerogel
does not change at arbitrary rotation in the orbital space, or continuous
degeneracy is preserved in a presence of the random field
$\eta_{jl}({\bf r})$. The order parameter
 $A_{\mu j}$ enters the tensor of superfluid densities in the combination
$A_{\mu l}A_{\mu j}^* + A_{\mu j}A_{\mu l}^*$ i.e. condition  (4) is the
requirement of the isotropy of that tensor.  It
might be well to point out here the difference between the argument of Imry and
Ma and the argument presented in this section. In the former case the
order parameter is assumed to be fixed by minimization of the free energy
without the random field. Possible
disordering is caused by a random walk of the order parameter over its space of
degeneracy. No other change of the order parameter  is admitted. In the latter
case all variations of the order parameter $A_{\mu j}$ are considered. That
makes possible an adjustment of the order parameter to the random field.
To emphasize the property of phases, which satisfy criterion (12) or (13) to
resist disordering action of the random field a term ``robust" was suggested
\cite{fom2}.

Hence, the procedure of finding of order parameters of possible superfluid
phases of liquid $^3$He in a presence of the random field $\eta_{jl}({\bf r})$
begins with a choice of a family of matrices $\bar A_{\mu j}$ meeting
criterion (4). These matrices form a "zero-order approximation" for the order
parameter.
As a next step the first order corrections $a_{\mu j}$ and $a^*_{\mu j}$ are
expressed in terms of $\bar A_{\mu j}$ and $\eta_{jl}({\bf r})$ via eqns.
(9) and (10). In a practice it is more convenient to work with the Fourier
transforms $a_{\mu j}({\bf k})$. After calculation of the averages
$<a_{\nu n}a_{\beta l}>$ etc. and substitution of these averages in eq.(8) it
 becomes a closed equation for $\bar A_{\mu j}$.
 Coefficients $\beta_1,...\beta_5, K$ and correlation functions $<\eta_{\nu
n}({\bf k})\eta_{\beta l}({-\bf k})>$ have to be given.
When $\eta_{jl}({\bf r})=0$ a conventional equation for the extrema of
free energy of pure $^3$He is recovered.

The order parameter of BW-phase
$$
   A^{BW}_{\mu j} = \Delta e^{i\varphi}R_{\mu j},     \eqno(15)
$$
where $R_{\mu j}$ -- a real orthogonal matrix, satisfies criterion (4).
If the dipole interaction is neglected this matrix up to a relative rotation
of spin space with respect to the orbital is equivalent to the unit matrix
$\delta_{\mu j}$. Aerogel is assumed to be uniform and isotropic. In that case
 tensor structure of correlation functions
$<\eta_{\nu n}({\bf k})\eta_{\beta l}({-\bf k})>$ is determined by symmetry
 \cite{fom}. It is clear that with that functions the order parameter
 proportional to the unit matrix satisfies eq. (8) and the BW-phase can be a
 minimum of free energy in a presence of aerogel. In comparison with the pure
$^3$He the values of phenomenological coefficients $\beta_1,.. \beta_5$  will
change. This influences a region of stability of BW-phase and changes
thermodynamic quantities, depending on these coefficients. We don`t consider
here a problem of explicit calculation of corrections to $\beta_1,... \beta_5$.

The order parameter of ABM-phase
$$
A_{\mu j}=\Delta\frac{1}{\sqrt{2}}\hat d_{\mu}(\hat m_j+i\hat n_j), \eqno(16)
$$
does not satisfy criterion (4). This brings up the question:
what kind of the order parameter can describe the properties of
A-like phase.

\section {Zero-order approximation for ESP phases}
The measured magnetic susceptibility of A-like phase coincides with that
of the normal phase \cite{osher}. It means that the order parameter of that
phase does not contain components corresponding to zero projection of the spin
of Cooper pair on the direction of magnetic field i.e. it belongs to ESP
(equal spin pairing) type. General form of the ESP-type order parameter is:
$$
A_{\mu j}=\Delta\frac{1}{\sqrt{3}}[\hat d_{\mu}( m_j+i n_j)+
\hat e_{\mu}( l_j+i p_j)]        , \eqno(17)
$$
where $\hat d_{\mu}$ and $\hat e_{\mu}$ are mutually orthogonal unit vectors,
$m_j,n_j,l_j,p_j$ -- arbitrary vectors.
Direct substitution of the order parameter (17) into eq. (4) shows that
the criterion is met if the vectors $m_j,n_j,l_j,p_j$ satisfy the equation
$$
m_jm_l+n_jn_l+l_jl_l+p_jp_l=\delta_{jl}.                 \eqno(18)
$$
The normalization condition $A_{\mu j}A_{\mu j}^*=\Delta^2$ is used.
One of the solutions of eq. (18):  ({\bf p}=0, {\bf m,n,l}-- orthonormal
basis) has been discussed before \cite{fom_A}.
To find all solutions consider four four-dimensional vectors
 $M_s, N_s, L_s, P_s$ (s=1,2,3,4) which obey the equation
$$
 M_rM_s+N_rN_s+L_rL_s+P_rP_s=\delta_{rs},                     \eqno(19)
$$
Unique (up to a general rotation and reflections) solution of this equation
is a set of four orthonormal vectors
$\hat q^{(a)}$ : $\hat q^{(a)}\cdot\hat q^{(b)}=\delta^{ab}$.
Let us take an arbitrary unit four-vector
$\hat\nu=(\nu_1,\nu_2,\nu_3,\nu_4)$ and take projections of  the vectors
$\hat q^{(a)}$ on the three-dimensional hyperplane orthogonal to
$\hat\nu$.
As a result we obtain four three-dimensional vectors:
$$
{\bf m}=\hat q^{(1)}-\nu_1\hat\nu, {\bf n}=\hat q^{(2)}-\nu_2\hat\nu,
{\bf l}=\hat q^{(3)}-\nu_3\hat\nu, {\bf p}=\hat q^{(4)}-\nu_4\hat\nu, \eqno(20)
$$
Taking a product of
$m_jm_l+n_jn_l+l_jl_l+p_jp_l$, on an arbitrary vector $a_l$ perpendicular to
$\hat\nu$, and using for {\bf m,n,l,p} expressions (20)
one can see that these vectors satisfy eq. (18).
With the aid of formulae (20) one can find other useful properties of the
vectors {\bf m,n,l,p}:
$$
  m^2+n^2+l^2+p^2=3                                       \eqno(21)
$$
$$
  {\bf m}\cdot{\bf n}=-\nu_1\nu_2, {\bf m}\cdot{\bf l}=-\nu_1\nu_3,
  {\bf n}\cdot{\bf l}=-\nu_2\nu_3, ...                     \eqno(22)
$$
$$
  m^2=1-\nu_1^2, n^2=1-\nu_2^2, ...                        \eqno(23)
$$
From eqns. (22) follows that
$[{\bf m}\times{\bf n}]\cdot [{\bf l}\times{\bf p}]=0 $,
i.e. the normals to the planes defined by the pairs of vectors
{\bf m,n} and {\bf l,p} are mutually orthogonal. This property is preserved
for arbitrary choice of pairs among the four vectors {\bf m,n,l,p}.
Hence eq. (17) with the vectors {\bf m,n,l,p}, defined according to eq. (20)
specifies a three-parameter family of quasi isotropic order parameters
of ESP-type. Substitution of this order parameters in eq. (5) renders
zero-order energies of the corresponding phases:
$$
\frac{F^{(0)}_{GL}}{N(0)}=\tau\Delta^2+\frac{\Delta^4}{18}
[\beta_1+9\beta_2+\beta_3+5(\beta_4+\beta_5)-
4(\beta_1+\beta_5)(\nu_1\nu_4-\nu_2\nu_3)^2].                      \eqno(24)
$$
Parameters $\nu_1,\nu_2,\nu_3,\nu_4$ enter this expression in a combination
 $\Lambda\equiv(\nu_1\nu_4-\nu_2\nu_3)$. If $(\beta_1+\beta_5)\equiv\beta_{15}<0$,
 the minimum of free energy corresponds to  $\Lambda=0$ i.e. to
$$
\nu_1\nu_4=\nu_2\nu_3.  \eqno(25)
$$
In a weak coupling approximation both $\beta_1$ and $\beta_5$ are negative
and the inequality $(\beta_1+\beta_5)<0$ is well satisfied.
 Condition (25) has simple physical meaning. The order parameters given by
 eq. (17) are generally speaking nonunitary. The corresponding phases
 can have a finite spin density, which is proportional to
  $e_{\mu\nu\lambda}A_{\mu j}A_{\nu j}^*$ or, explicitly to
$(2\Delta^2/3)[\hat d\times\hat e][{\bf n}\cdot{\bf l}-{\bf m}\cdot{\bf p}]$.
 With the aid of eq. (22) one can find that the spin density vanishes if
 eq. (25) is satisfied. This condition specifies two-parametric family of
 nonferromagnetic quasi isotropic phases which possibly includes A-like phase.
 A possible parametrization of this family is:
$\nu_1=\sin\alpha\sin\beta, \nu_2=\sin\alpha\cos\beta,
\nu_3=\cos\alpha\sin\beta, \nu_4=\cos\alpha\cos\beta$.
The most symmetrical nonferromagnetic phase corresponds to the choice
$\alpha=\pi/4,\beta=\pi/4$. With these values of parameters
$\nu_1=\nu_2=\nu_3=\nu_4=1/2$ and the absolute values of vectors
{\bf m,n,l,p} are all equal to  $\sqrt{3}/2$. The angles between every two of
the four vectors {\bf m,n,l,p} are also equal. Such set is formed by vectors
 connecting the center of a  right tetrahedron with its vertices.

If $\beta_{15}>0$ then $\Lambda^2$ in equilibrium reaches its maximum value
 1/4 which is reached at  $\nu_1=\nu_4 ;\nu_2=-\nu_3$ or at
$\nu_1=-\nu_4 ;\nu_2=\nu_3$. In both cases it is the one-parametric family.
In the first case it can be parmetrised as
 $\nu_1=\nu_4=\frac{1}{\sqrt{2}}\sin\gamma$
$\nu_2=-\nu_3=\frac{1}{\sqrt{2}}\cos\gamma$. The most symmetrical ferromagnetic
phase  corresponds to
$\nu_1=-1/2, \nu_2=\nu_3=\nu_4=1/2$  i.e. it is obtained from the most
symmetrical nonferromagnetic solution by inversion of one of the vectors
 $m_j,n_j,l_j,p_j$.

\section {Effect of magnetic field}
In a magnetic field two terms have to be added to free energy. One is
quadratic in the field:
$$
f_H^{(2)}=-\frac{1}{2}\chi_{\mu\nu}H_{\mu}H_{\nu}.           \eqno(26)
$$
For all ESP phases one of the principal values of a tensor of magnetic
susceptibility $\chi_{\mu\nu}$ coincides with the susceptibility of the normal
phase $\chi_n$. In a vicinity of $T_c$ tensor $\chi_{\mu\nu}$ is determined by
symmetry argument $\chi_{\mu\nu}=\chi_n\delta_{\mu\nu}- \kappa(A_{\mu
j}A_{\nu j}^*+A_{\nu j}A_{\mu j}^*)$. Second term in the r.h.s. describes
a decrease of the transverse susceptibility in comparison with the normal
value $\chi_n$. This is  two-dimensional tensor with the principal values
$\frac{2\Delta^2}{3}\lambda_{1,2}$, where $\lambda_{1,2}$ are the roots of
the equation
$$
\lambda^2-3\lambda+2+\Lambda^2=0.
$$
In a nonferromagnetic phase  $\Lambda=0$, $\lambda_1=2$, $\lambda_2=1$ i.e.
transverse susceptibility is anisotropic.  In a ferromagnetic phase
$\Lambda^2=1/4$, $\lambda_1=\lambda_2=1$, and transverse susceptibility is
 isotropic. Equilibrium orientation of the order parameter corresponds to
 maximum susceptibility $\chi_{\mu\nu}$ in the direction of magnetic field
 and additional energy eq. (26) is the same for all ESP-phases.

Except for the quadratic there is a linear on magnetic field term in the
free energy:
$$
f_H^{(1)}=i\zeta e_{\mu\nu\lambda}A_{\mu j}A_{\nu j}^*H_{\lambda}.  \eqno(27)
$$
In the bulk $^3$He this term gives rise to the splitting of the transition
from the normal phase in the A-phase in two transition with close
temperatures. Ferromagnetic A$_1$-phase appears first, it contains Cooper
pairs with only one spin projection. At a lower temperature transition in
A$_2$-phase occurs with both spin projections present in the condensate.
The coefficient $\zeta$ is proportional to the derivative of the density of
states over energy and the temperature interval where  A$_1$-phase exists
is small as $\mu H/\varepsilon_F$, where $\mu$ is the magnetic moment of
$^3$He nucleus and $\varepsilon_F$  - Fermi energy.

Let us take into account additional free energy eq. (27) in a presence of
aerogel.
$$
\frac{F^{(0)}_{GL}}{N(0)}=(\tau-\frac{\zeta H\Lambda}{3})\Delta^2-
\frac{2\Delta^4}{9}\beta_{15}\Lambda^2+\frac{\Delta^4}{18}
[\beta_1+9\beta_2+\beta_3+5(\beta_4+\beta_5)].                   \eqno(28)
$$
This expression has to be minimized over $\Lambda$ and over $\Delta^2$.
The result of minimization depends on a sign of
$\beta_{15}$. If $\beta_{15}>0$, then for all $\Delta^2$ the energy reaches its
minimum at $|\Lambda|=1/2$, i.e. the ferromagnetic phase is stable.
Transition in the superfluid state takes place at
$\tau=\zeta H/6$. At $\tau<\zeta H/6$  $\Delta^2=-9\tau_H/B$, where
$\tau_H=\tau-\zeta H/6$, $B=9\beta_2+\beta_3+5\beta_4+4\beta_5$.
Magnetic moment has a field independent increment which is proportional to
 $\Delta^2$: $M=N(0)\zeta\Delta^2/6$.

If $\beta_{15}<0$, the ferromagnetic phase minimizes free energy (28) only
in the temperature interval
$[1+(B/\beta_{15})](\zeta H/6)<\tau<\zeta H/6$.
At  $\tau_2=[1+(B/\beta_{15})](\zeta H/6)$ a second transition takes place in
the phase with  $\Lambda=-\frac{3\zeta H}{4\beta_{15}\Delta^2}$.
When temperature is lowered $\Lambda\to 0$ i.e. an additional magnetic
moment vanishes. The transition at $\tau=\tau_2$ is analogous to  $A_1\to A_2$
transition in bulk $^3$He. Hence the considered ferromagnetic phases are
analogous to A$_1$-phase of bulk $^3$He except that for the new phases
the pairing amplitude is finite for both projections of spin $s=1$ and $s=-1$.

\section {Discussion}
We conclude that in $^3$He  in a presence of aerogel only such superfluid
phases can exist  for which the order parameter satisfies criterion (4).
The examples of A-like and A$_1$-like phases demonstrate that this condition
does not  determine a matrix  $\bar A_{\mu j}$ uniquely, but selects a family
of such matrices. For the further selection a next approximation on the random
field $\eta_{jl}({\bf r})$ has to be considered. The answer depends on
correlation functions of the random field, which are not known.
A class of admitted solutions can be restricted with the aid of the observed
 properties of the phases. For example the splitting of the phase transition in
  a magnetic field is the evidence that $\beta_{15}<0$ and away from the $T_c$
 nonerromagnetic phase is stable. If the $A_1\to A_2$  transition is absent
 then a ferromagnetic phase is stable ($\beta_{15}>0$). Experimental situation
 in that respect is not quite clear.

 A characteristic property of all quasi isotropic phases, which can be used
 for a general check of the proposed scheme is the isotropy in a principal
 order on $\eta_{jl}({\bf r})$ of the tensor of superfluid densities.
 In particular this is a difference between the proposed A-like phases and
 of the A-phase of bulk  $^3$He.

 \section {Acknowledgments}

I acknowledge V.V.Dmitriev and J.Parpia for usful discussions and comments
E.I.Kats for the invitation to ILL in Grenoble where part of this work
has been done and for stimulating discussions. This work was supported
by RFBR-foundation, grant 01-02-16714 and by Ministry of Industry, Science and
Technologies of Russia.

  \end{document}